\documentclass[%
reprint,
superscriptaddress,
groupedaddress,
showpacs,
preprintnumbers,
nofootinbib,
 amsmath,amssymb,
 aps,
 prl,
]{revtex4-1}

\usepackage{graphicx}
\usepackage{dcolumn}
\usepackage{bm}
\usepackage[utf8]{inputenc}
\usepackage{textcomp}
\usepackage[thickqspace, squaren]{SIunits} 
\usepackage{color}
\usepackage{braket}
\usepackage[hidelinks,pdfstartpage=1]{hyperref}
\hypersetup{pdfstartview={XYZ null null 1.2}}

\newcommand{\micron}{\ensuremath{\mu {\rm m}} }

\newcommand{\rig}{\Ket{R}}
\newcommand{\lef}{\Ket{L}}
\newcommand{\eigone}{\Ket{1}}
\newcommand{\eigtwo}{\Ket{2}}

\begin{document}


\title{A microwave chip-based beam splitter for low-energy guided electrons}
\author{Jakob Hammer}
\email{jakob.hammer@fau.de}
\author{Sebastian Thomas}
\author{Philipp Weber}
\author{Peter Hommelhoff}
\affiliation{Department für Physik, Friedrich-Alexander-Universit\"at Erlangen-N\"urnberg, Staudtstr. 1~\mbox{, 91058~Erlangen,  Germany}}
\date{\today}

\begin{abstract}   
We demonstrate the splitting of a low-energy electron beam by means of a microwave pseudopotential formed above a planar chip substrate. Beam splitting arises from smoothly transforming the transverse guiding potential for an electron beam from a single-well harmonic confinement into a double well, thereby generating two separated output beams with $5\,$mm lateral spacing. Efficient beam splitting is observed for electron kinetic energies up to $3\,$eV, in excellent agreement with particle tracking simulations. We discuss prospects of this novel beam splitter approach for electron-based quantum matter-wave optics experiments.
\end{abstract}

\maketitle
A beam splitter is the quintessential component in many modern physics experiments. The visualization of the quantum mechanical phase hinges on it. Its various realizations have enabled the observation of fundamental physics phenomena such as quantum optics experiments with photons~\cite{Mandel1995}, many-body interference experiments with cold atoms in optical lattices~\cite{Bloch2008}, neutron interferometry~\cite{Rauch2000} and fundamental interference studies with heavy molecules~\cite{Juffmann2012}. Prominent among these studies are interference experiments with electrons, which have enabled groundbreaking insight into, for example, the wave-particle duality with massive particles~\cite{Davisson1927,Boersch1943,Marton1953,Tonomura1989} and the Aharanov-Bohm effect~\cite{Tonomura1986}.

A plethora of electron interferometry experiments~\cite{Hasselbach2010} was triggered by the invention of the electrostatic biprism in 1955~\cite{Mollenstedt1955}. It is a relatively rugged transverse beam splitting element that also serves as a workhorse in modern commercial electron microscopes employing holographic techniques~\cite{Gabor1948,Tonomura1999}. In particular, interference experiments with low-energy electrons have demonstrated reduced radiation damage allowing the nondestructive imaging of biological molecules~\cite{Germann2010}.

An entirely new electron optical toolkit arises from the manipulation of slow electrons in free space using a microwave quadrupole guide~\cite{Hoffrogge2011}. The generation of the necessary high-frequency electric fields by means of a planar microwave chip provides ease of scalability and the flexibility to engineer versatile guiding potentials in the near-field of the microwave excitation. This renders surface-electrode structures ideally suited for the implementation of electron beam splitters or resonators with prospects for novel quantum optics experiments with guided electrons. Based on a similar technology, surface-electrode ion traps have been employed to provide finely structured potential landscapes. For example, junctions for trapped ions have been realized~\cite{Amini2010,Moehring2011,Shu2014,Pearson2006,Wright2013,Hensinger2006}, or double-well potentials with small distances between the potential minima to couple separately trapped ions via the Coulomb force~\cite{Brown2011, Harlander2011}. In this letter we show the concept and the experimental demonstration of a new beam splitter for guided electrons with kinetic energies in the electron-volt range.

Oscillating electric fields allow the generation of a time-averaged restoring force to confine the motion of charged particles in free space~\cite{Major2005}. The microwave guide for electrons is based on a two-dimensional, high-frequency quadrupole potential providing transverse confinement, similar to a linear Paul trap~\cite{Paul1990}. The details of this concept are summarized in the Supplemental Material. Stable operation of the guide practically requires oscillation frequencies of the microwave drive in the gigahertz range. The resulting tight transverse confinement is described by a time-averaged, harmonic pseudopotential. Moreover, electrons can be confined in the saddle point of any inhomogeneous high frequency electric potential $\phi(\vec{r},t)=\phi_{RF}(\vec{r})\cos(\Omega t)$ with drive frequency $\Omega$ if the potential gradient is nearly constant over the range of the electron's oscillation~\cite{Major2005}. We generate such an electric potential by means of a planar microwave chip. As a key feature, this chip-based technology provides the unique possibility to achieve high field gradients in the near-field of a microstructured electrode design allowing for precise control over the motion of the guided electrons.

\begin{figure}[hbt]
  \centerline{\includegraphics[width=8.6cm]{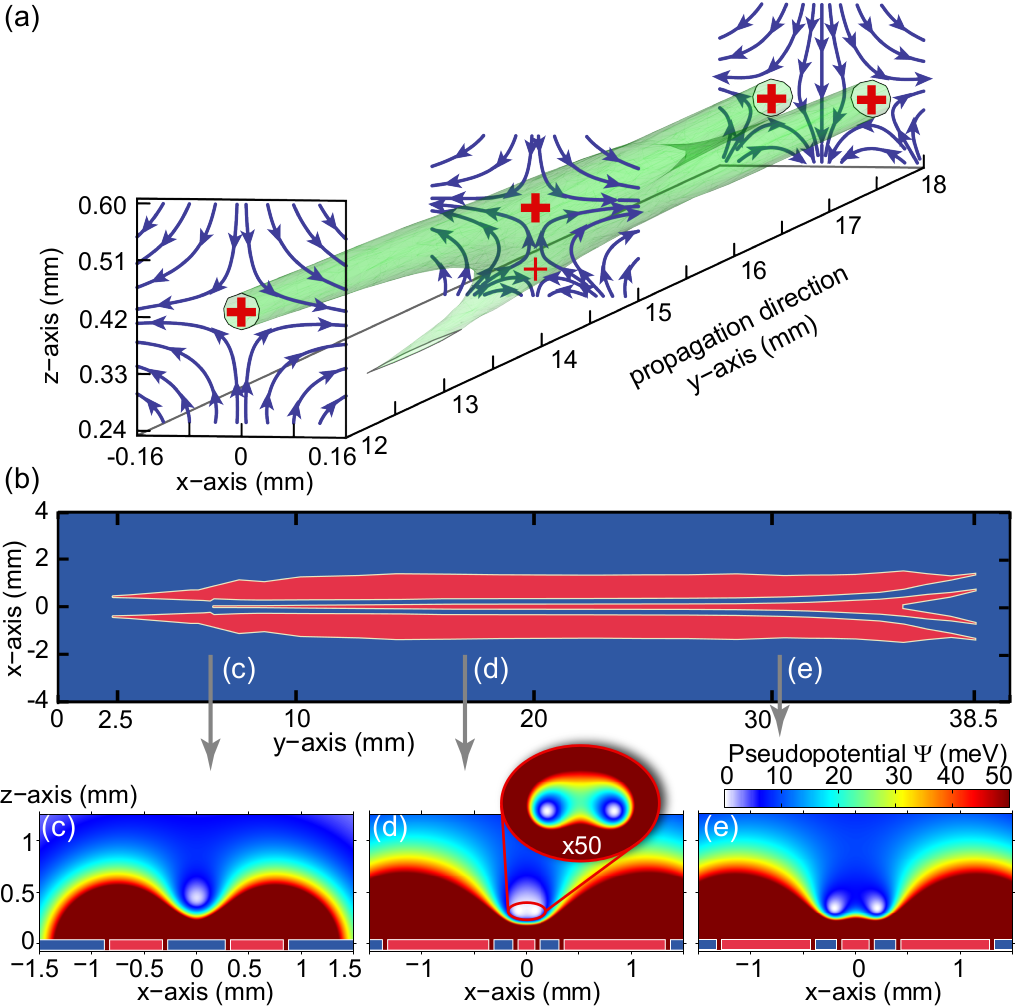}}
  \caption[guidingPot]{(Color online) Electrode design of the planar beam splitter chip and pseudopotential simulations. (a) Electric field line plots together with an isopotential surface of the guiding potential at $0.25\,$meV.   (b) Numerically optimized chip electrodes with microwave signal applied to the red electrodes. The remaining blue area is grounded. By means of the tapered central electrode the transition from a quadrupole to a hexapole electric field symmetry is achieved, as shown in (a). (c) Cut through the electrode plane at $y=6.5\,$mm showing the simulated pseudopotential in the transverse plane. The pseudopotential minimum forms at a height of $450\,\micron$ above the substrate providing harmonic confinement. (d) At  $y=17$\,mm the additional central electrode, with a width of $160\,\micron$, results in the formation of a double-well pseudopotential with a separation of $150\,\mu$m between the minima. A fourfold magnified zoom-in is shown in the inset with a 50 times amplified color code. By increasing the width of the center electrode the separation of the double-well minima is gradually increased. (e) At $y=30$\,mm  the central electrode is $260\,\micron$ wide, leading to a separation of the minima of $400\,\mu$m.
  }
  \label{fig:splitterIntro}
\end{figure}
For the on-chip splitting of the guided electron beam we incorporate a junction in the guiding potential by gradually transforming the driving electric field from a quadrupole to a hexapole symmetry along the chip. Using hexapole electric fields a junction can be realized in the pseudopotential~\cite{Wesenberg2009}. Figure~\ref{fig:splitterIntro}(a) illustrates electric field line plots in the transverse \textit{xz}-plane at three locations along the planar electrode structure. Additionally an isopotential surface of the guiding potential at $0.25\,$meV is shown, with microwave drive parameters as given below. The electric field line plots and the isopotential surface plot have been obtained by simulating the electric field that is created by the surface electrodes, the design of which is shown in Fig.~\ref{fig:splitterIntro}(b)~\cite{Hommelhoff2014}. The microwave signal is applied to the red electrodes, whereas the ground plane is indicated in blue. At a position of $y=12\,$mm along the chip, the electric field in the transverse plane is governed by a strong quadrupole component leading to the creation of a saddle point guiding electrons in the center, as indicated by the red cross. By changing the width of the tapered signal electrode in the center, the electric field above the guiding chip can be transformed along the $y$-direction from a quadrupole to a hexapole symmetry. The hexapole field component gives rise to an additional saddle point that continuously approaches the guiding potential minimum from the chip surface. This is indicated in the field line plot at $y=15\,$mm, where two saddle points form on the vertical $z$-axis. Further along the chip, for increasing $y$, both saddle points merge in the $xz$-plane and subsequently separate in the transverse $x$-direction.

An electric field with a predominant quadrupole component may be generated by five electrodes on a planar chip substrate~\cite{Wesenberg2008}. Figure~\ref{fig:splitterIntro}(c) shows a cut through the electrode structure at $y=6.5\,$mm together with a simulation of the pseudopotential in the $xz$-plane. As a result of the strong quadrupole component, a single guiding potential minimum forms at a height of $450\,\micron$ above the chip surface. The simulation is performed with a microwave drive frequency $\Omega=2\pi\cdot990\,$MHz and a voltage amplitude $V_0=16\,$V on the signal electrodes. Figure~\ref{fig:splitterIntro}(d) shows a cut through the electrode plane further along the chip at $y=17\,$mm. Here it comprises seven electrodes with a microwave signal electrode in the center. This leads to the creation of a strong hexapole field component giving rise to a double well in the pseudopotential. By adjusting the width of the central electrode, the separation of the double-well minima can be controlled. The distance between them is $150\,\micron$ in Fig.~\ref{fig:splitterIntro}(d) and $400\,\micron$ in Fig.~\ref{fig:splitterIntro}(e), which shows the simulated pseudopotential at $y=30\,$mm. The barrier height between the wells is $0.5\,$meV at $y=17\,$mm and $11.5\,$meV at $y=30\,$mm.

We have numerically optimized the electrode layout of the microwave chip using the \textit{Surface Pattern} package~\cite{NoteSurfacePattern,Schmied2010,Schmied2009}. The hexapole symmetry of the electric field close to the intersection point results in a junction with two incoming and two outgoing channels. By means of a systematic variation of the shape of the chip electrodes, we have reduced distortions in the beam splitter potential that arise from the additional incoming channel and minimized its impact on the trajectories of guided electrons. Details are given in the Supplemental Material. 

The microwave signal is delivered to the signal electrodes [drawn in red in Fig.~\ref{fig:splitterIntro}(a)] by a coplanar waveguide structure on the backside of the chip (not shown), which is interconnected to the top side by laser-machined, plated through-holes (see the Supplemental Material for details). The experiments are performed with $\Omega=2\pi\cdot990\,$MHz and an on-chip microwave power of $4.3\,$W, which results in $V_0\approx16\,$V~\cite{NoteElectricallyShort}.

A home-built thermionic electron gun~\cite{Erdman1982} provides an electron beam with kinetic energies down to $1\,$eV and beam currents on the order of several ten femtoamperes. As a result of this low electron current electron-electron interaction effects are irrelevant. The beam is collimated using two apertures resulting in a full opening angle of $14\,$mrad and a spot diameter of about $100\,\micron$ at the guide entrance. Behind the microwave chip electrons are detected on a microchannel plate (MCP) electron detector~\cite{NoteMCP} after traveling $10\,$mm in free space. Images of the phosphor screen behind the MCP are recorded by a CCD camera~\cite{NoteCCD}.

\begin{figure}[bt]
  \centerline{\includegraphics[width=8.6cm]{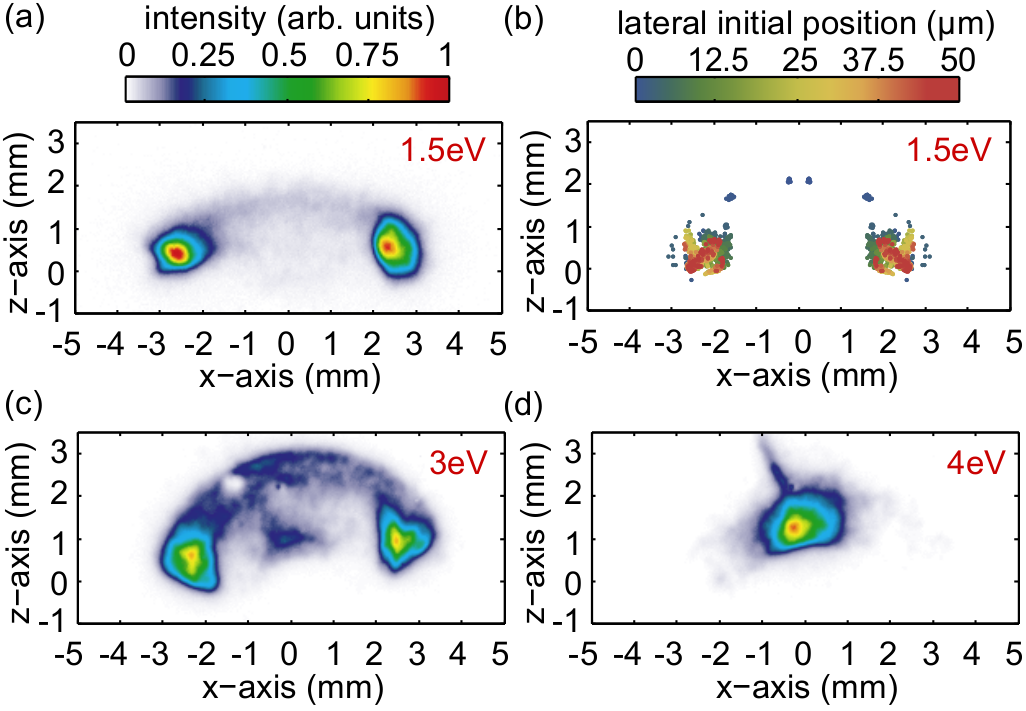}}
  \caption[guidingPot]{(Color online) Experimental (a) and simulated (b) detector signal of the split electron beam with $E_\mathrm{{kin}}=1.5\,$eV. (a) Clearly two guided beams are visible containing $80\%$ of all detected electrons. A faint signal of lost electrons is detected between the guided electron spots (between $x=-1.5$\,mm and $1.5$\,mm). The color scale depicts the intensity of the raw CCD image. (b) Simulated beam splitter signal based on trajectory simulations.  All signatures including the position and size of the output beams and the electron loss are reproduced by the simulation. The color scale corresponds to the initial lateral displacement of the electron trajectories along \textit{x}. See text for details. The dependence of the detected electron signal on the kinetic energy is shown for $3$\,eV (c) and $4$\,eV (d). For $4$\,eV the beam splitting potential is too weak to split up the beam.
  }
  \label{fig:splitterExp}
\end{figure}

Fig.~\ref{fig:splitterExp}(a) shows the detector signal recorded for an electron kinetic energy of $1.5\,$eV and the microwave parameters given above. We observe an electron signal with two symmetrically split up components. The distance between the two main spots is $5\,$mm, whereas each spot has an average full-width at half-maximum diameter of $0.75$\,mm. Additionally a faint signal of lost electrons is detected between the two guided components. The guided electrons comprise $80\%$ of the detected signal. Clearly, the injected electron beam is split into two collimated output beams.

In order to fully understand the observed features we perform classical particle tracking simulations. We release electron trajectories from a disk with a diameter of $100\,\micron$ and propagate them numerically in the simulated electric field of the beam splitter chip. Fig.~\ref{fig:splitterExp}(b) shows the resulting simulated electron signal, which is in excellent agreement with the experimentally observed output signal. The color scale illustrates the initial lateral displacement of the electrons along the \textit{x}-axis. Evidently, electrons released closest to the symmetry axis of the beam splitter potential [blue dots in Fig.~\ref{fig:splitterExp}(b)] are preferentially lost. This can be understood by considering the extreme case of an electron being released at $x=0\,$mm. Because of the planar symmetry of the beam splitter potential in the $x$-direction, such a classical trajectory does not encounter any transverse potential gradient and therefore no deflecting force along $x$. As a result, this trajectory cannot follow the pseudopotential minimum paths of the separating double well and is only deflected vertically away from the substrate. For this reason, electrons that propagate closest to the symmetry axis may preferentially become lost from the beam splitter potential. Using quantum mechanical simulations we show in the Supplemental Material that \textit{lossless}, adiabatic splitting of an electron beam can be achieved by means of an optimized beam splitter potential.

Further, we have varied the electron kinetic energy from $1.5\,$eV to $3\,$eV. We find that the signal of lost electrons becomes larger with energy as depicted in Fig.~\ref{fig:splitterExp}(c) as compared to Fig.~\ref{fig:splitterExp}(a). This is because with increasing forward momentum of the electrons the transverse gradient of the beam splitter potential becomes insufficient to significantly deflect the electrons in the lateral \textit{x}-direction. Accordingly, the electron trajectories cannot follow the separating paths of the potential minimum and are lost from the potential. As a consequence, for energies above $4\,$eV we observe no splitting anymore and all electrons are detected around $x=0\,$mm in Fig.~\ref{fig:splitterExp}(d).

The beam diameter of $100\,\micron$, attained with the thermionic electron gun, is not matched to the diameter of the quantum mechanical ground state wavefunction (on the order of $100\,$nm) of the transverse beam splitter potential. As a result, we estimate that the guided electrons fill up the potential up to energies of $0.75\,$meV in the current experiment, which is orders of magnitude larger than the quantum ground state energy on the order of $0.1\,\mu$eV. Therefore, the experiment is well described by classical particle tracking simulations. However, the direct injection of electrons into low-lying motional quantum states should be possible by matching the incoming electron beam to the ground state wavefunction of the transverse guiding potential~\cite{Hammer2014}. 

Ultimately, the wave-optical propagation of a guided electron is governed by discretized motional quantum states of the transverse guiding potential.  In the following, we illustrate the properties of the microwave beam splitter quantum mechanically and discuss prospects for electron-based quantum optics experiments. 

\begin{figure}[hbt]
        \centerline{\includegraphics[width=8.6cm]{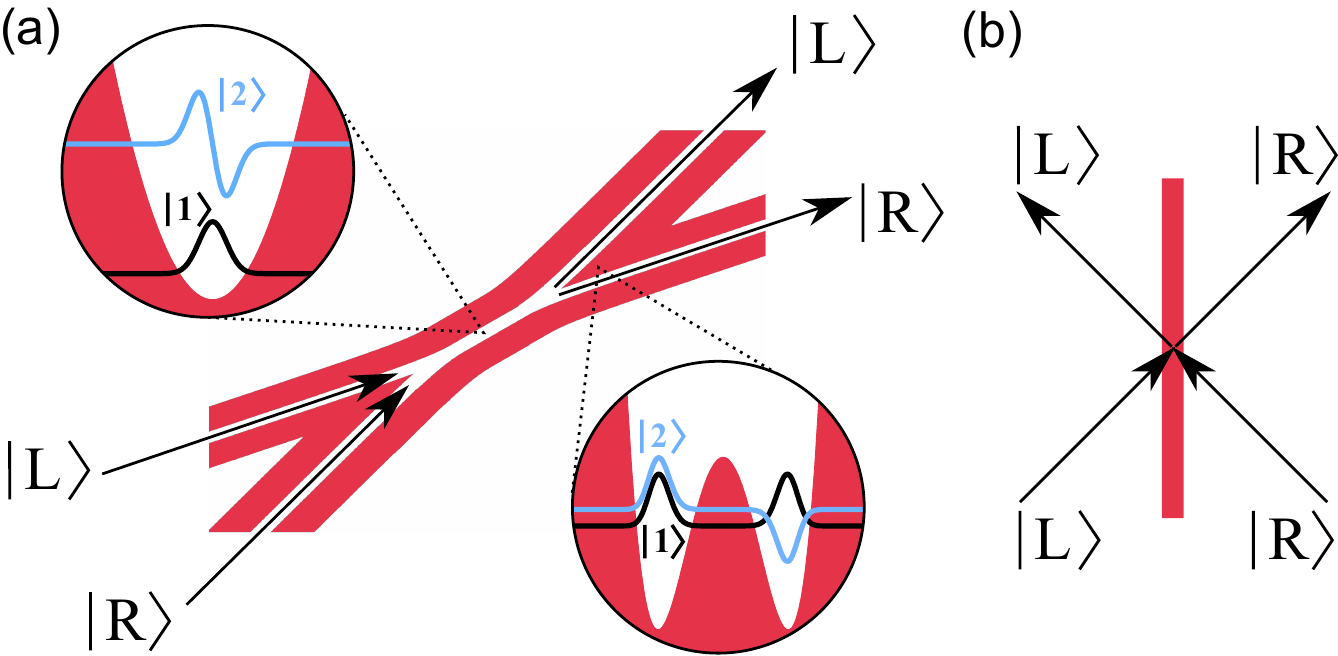}}
  \caption[bssketch]{(Color online) Sketch of an adiabatic microwave beam splitter (a) in comparison to a typical amplitude beam splitter as used in light optics (b).
  }
  \label{fig:mwbs}
\end{figure}

It is instructive to compare the microwave beam splitter for electrons to a typical \textit{amplitude} beam splitter as used in light optics. As detailed above, the beam splitter potential based on a hexapole intersection features two incoming and two outgoing channels. For simplicity we consider a planar symmetry of the beam splitter potential around the intersection point along $y$, as indicated in Fig.~\ref{fig:mwbs}(a). We label an incoming electron that occupies the motional ground state of the left (right) arm of the beam splitter with the state $\lef$ ($\rig$). To understand the evolution of these localized input states one needs to consider the transverse energy eigenstates $\eigone$ and $\eigtwo$ at different points along the length of the beam splitter [see the insets of Fig.~\ref{fig:mwbs}(a)]. While the paths are spatially well separated by a potential barrier these are the symmetric and antisymmetric ground states of a double-well potential, and their energy is (almost) degenerate. The localized input states are a superposition $\lef=(\eigone+\eigtwo)/\sqrt{2}$ and $\rig=(\eigone-\eigtwo)/\sqrt{2}$ of these eigenstates.

As $\lef$ and $\rig$ are not energy eigenstates, electrons will in principle tunnel between these two states. However, as long as the potential barrier is much larger than the transverse energy of these states, both wells are separated and the tunneling frequency is negligibly low. From a quantum mechanical point of view, the function of the beam splitter is to increase this frequency by bringing the two wells closer together and eventually merging them. In the center part of the splitter, the superposition states $\lef$ and $\rig$ are then no longer spatially separated and, hence, wave amplitude is transferred between $\lef$ and $\rig$. In general an incoming state with amplitudes $l$ in the left and $r$ in the right path is turned to an outgoing state with amplitudes $l'$ and $r'$. If we describe the left path by the state $\lef=\bigl(\begin{smallmatrix}1\\0\end{smallmatrix}\bigr)$ and the right path by the state $\rig=\bigl(\begin{smallmatrix}0\\1\end{smallmatrix}\bigr)$, the effect of the beam splitter $B$ can be described as a multiplication of the state with a unitary matrix: $\bigl(\begin{smallmatrix}l'\\r'\end{smallmatrix}\bigr)=B\bigl(\begin{smallmatrix}l\\r\end{smallmatrix}\bigr)$. If we disregard phase shifts, $B$ is essentially a rotation matrix whose angle depends on the oscillation frequency $\omega$ between $\lef$ and $\rig$ and the time the electron spends in the center part of the splitter. 

The previous discussion assumes that the electron initially occupies the motional ground state of the transverse guiding potential. As described above, this can be achieved using a diffraction-limited electron gun in order to match the injected electron beam to the ground state wavefunction of the guiding potential. Interestingly, a multi-mode interferometer using higher vibrational states has been investigated in the context of guided atom interferometry~\cite{Andersson2002}. Furthermore, the above description requires that an electron initially prepared in the quantum ground state maintains its state while propagating along the beam splitter. The current design lacks this crucial feature of adiabaticity. Using quantum mechanical simulations we have investigated the key prerequisites to achieve adiabatic splitting of the ground state mode. 
The details of the simulations are described in the Supplemental Material. The amount of transverse vibrational excitations depends critically on the geometric opening angle between the beam splitter paths as well as the energy separation of the transverse eigenstates. As one would expect, a smooth splitting process and, hence, a small opening angle is beneficial. By scaling the guiding potential transversely, we find that the half opening angle of the current design has to be reduced from 40\,mrad to $0.1\,$mrad. In addition, we have to increase the microwave drive frequency to $\Omega=2\pi\cdot8$\,GHz to obtain beam splitting with $90\%$ of the population remaining in the ground state mode after the splitting. The eightfold higher $\Omega$ effectively increases the curvature of the transverse potential and results in an $\sqrt{8}$-fold larger energy level separation of the single-well potential of $\Delta E\sim0.24\,\mu$eV and, hence, an oscillation frequency $\omega=\Delta E/\hbar\sim2\pi\cdot58\,$MHz. 
Both, the small beam splitter angle and the higher $\Omega$ require a re-design of the current microwave chip.

As just introduced, beam splitters used in quantum optics experiments [like in Fig.~\ref{fig:mwbs}(b)] are usually described by unitary matrices, which reflect the coupling between the amplitudes of two states~\cite{Zeilinger1981,Schleich2005}. The microwave beam splitter demonstrated here is a promising new technology because it may become such an \textit{amplitude} beam splitter for electrons. Most current experiments on electron interference rely on the electrostatic biprism, which is a \textit{wavefront} beam splitter. The wavefront beam splitter can be regarded as an electron optical device that generates two virtual sources by a spatial division of the beam. In that case, interference between both output beams relies on the spatial coherence of the electron source~\cite{Born1999}. In contrast, using an amplitude splitter the phase between both output beams and their amplitudes are fully determined by the physical properties of the beam splitter device. To this end, the manipulation of electrons using the microwave beam splitter augments the already available, rich electron optical toolkit and may herald new quantum optics experiments with free electrons. In particular, a novel quantum electron microscopy concept is emerging that employs multiple \textit{amplitude} splittings of a quantum particle's wavefunction for the noninvasive imaging of biological samples~\cite{Putnam2009, Thomas2014}.
\begin{acknowledgments}
We thank J. Hoffrogge, J. McNeur, P. Kruit and the QEM collaboration for discussions. This research is funded by the Gordon and Betty Moore Foundation.
\end{acknowledgments}
%



\clearpage
\begin{center}
\textbf{\large Supplemental Material}
\end{center}

\section{Microwave quadrupole guide for elecrons}
\label{sec:MWGuide}
For the guiding of electrons above the surface of a microwave chip we use a two-dimensional, high-frequency electric quadrupole potential $\phi(\vec{r},t)=\phi_{RF}(\vec{r})\cos(\Omega t)$, which provides a transverse harmonic pseudopotential to confine electrons along the guide's axis. In order to achieve stable confinement of electrons in the microwave guide, the frequency and the amplitude of the time-dependent electric quadrupole potential have to be matched to the electron's charge-to-mass ratio $Q/M$ and the spatial dimensions of the electrode structure generating the potential. The requirements on the microwave drive parameters can be obtained from the expression of a dimensionless stability parameter $q =\eta (Q/M) (2 V_0)/(\Omega^2R_0^2)$, where stable confinement of an electron requires $0<q<0.9$~\cite{Major2005}.  Here $V_0$ is the voltage amplitude applied to the electrodes and $R_0$ the height of the saddle point of the quadrupole potential above the chip surface, i.e. the position of the guide's center. Because of the high charge-to-mass ratio of electrons, the stable confinement in the quadrupole guide usually requires drive frequencies in the gigahertz range. Effectively, for small $q$, a time-averaged pseudopotential is generated by the oscillating electric potential, which is defined by $\Psi=Q^2/(4M\Omega^2)\,\left|\nabla \phi_{RF}(\vec{r})\right|^2$. The dynamics of an electron within the pseudopotential are then governed by an oscillatory macromotion with a frequency $\omega=(q/\sqrt{8})\,\Omega$ and a potential depth $U=(\eta/u)(q/8) V_0$. The constants $\eta$ and $u$ depend on the geometry of the planar electrode design~\cite{Wesenberg2008}.%

\section{Numerical optimization of the electrode layout}
\label{sec:NumOpti}
We have used the \textit{Surface Pattern} package~\cite{NoteSurfacePattern,Schmied2010,Schmied2009} to numerically optimize the shape of the chip electrodes. This package is implemented in Mathematica and is capable of analytically solving the Laplace equation of an arbitrary two-dimensional electrode structure in the gapless plane approximation. The optimization routine uses a Nelder-Mead simplex algorithm, which is a built-in function in Mathematica, to minimize a scalar merit function $M$ by systematic variation of the position of a predetermined number of points, which parametrize the shape of the chip electrodes. For the electrode optimization we have chosen $M$ to minimize vertical pseudopotential gradients $\partial\Psi/\partial z$ while maintaining a constant trap frequency $\omega_z$ in the vertical direction along the beam splitter path. A planar symmetry with respect to the $x=0$ plane is requested.

\begin{figure*}[hbt]
\centering\includegraphics[width=1\textwidth]{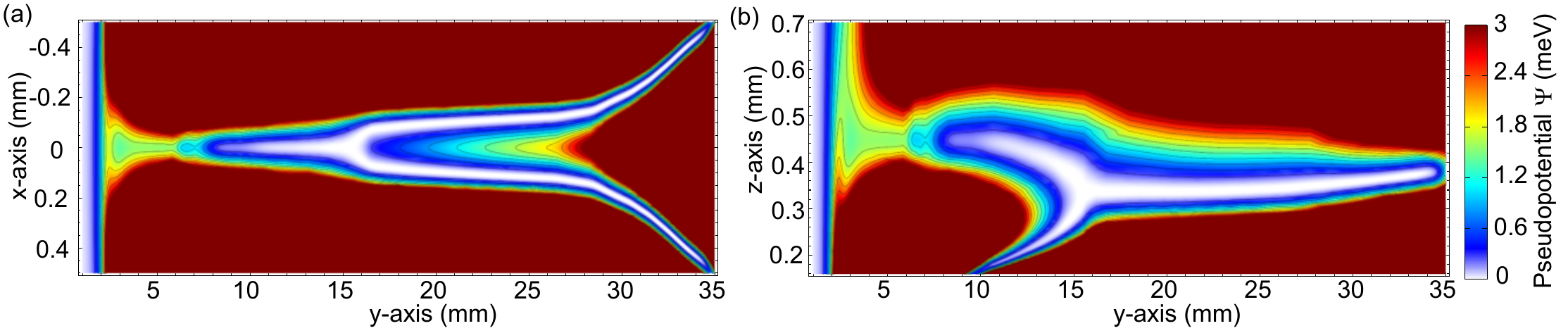}
\caption{Simulated pseudopotential $\Psi$ of the optimized beam splitter design. (a), Color plot of $\Psi$ in the $xy$-plane (birdseye-view on the chip surface). As the height of the pseudopotential minimum $z_{min}$ varies along the chip electrodes this plot is obtained by calculating $z_{min}$ for every point along $y$ and then plotting $\Psi(x,y,z_{min}(y))$. (b) Color plot of $\Psi$ in the vertical $zy$-plane. Here $x_{min}(y)$ is inserted for every position along $y$. The drive parameters are $\Omega=2\pi\cdot 1\,$GHz and $V_0=16\,V$.
\label{fig:splitterOptPseudoPot} }
\end{figure*}

\begin{figure*}[htb]
\centering\includegraphics[width=1\textwidth]{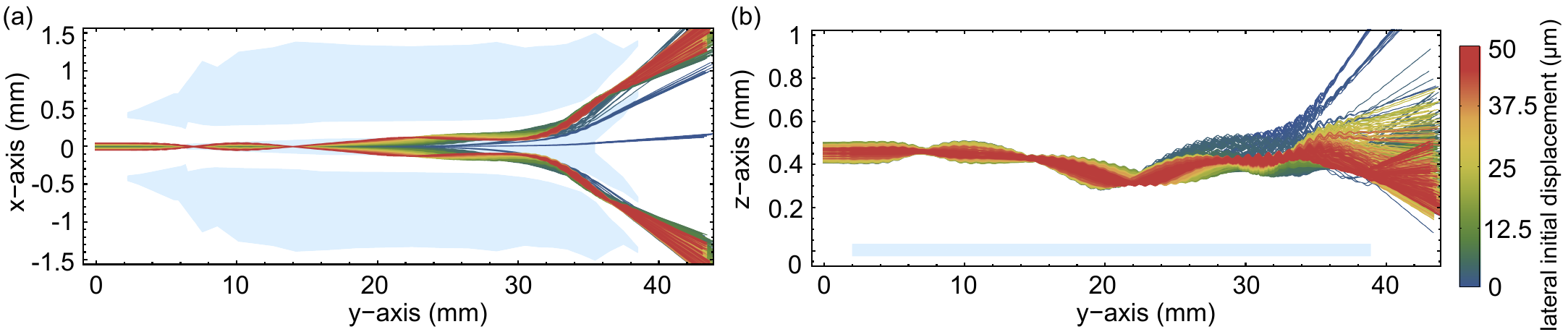}
\caption{Particle tracking simulations with $E_{kin}=1.5\,$eV. (a) Simulated electron trajectories in the $xy$-plane. The electrons perform a secular oscillation on the first $16\,$mm along the guide. At the beam splitter junction the beam becomes symmetrically divided and two split-up, guided beams are obtained. (b), Electron trajectories in the vertical $zy$-plane. Only trajectories that are released closest to the symmetry axis (blue lines) become lost in the vertical $z$-direction. The trajectories are simulated with $\Omega=2\pi\cdot1\,$GHz and $V_0=16\,$V.
\label{fig:splitterTraj} }
\end{figure*}

Fig.~1(a) shows the numerically optimized electrode design of the planar microwave chip. We can calculate the electric field created by this electrode structure and, hence, the pseudopotential $\Psi$ (according to the expression given above). The microwave drive parameters in the simulation are $\Omega=2\pi\cdot 1\,$GHz and $V_0=16\,V$. In Fig.~\ref{fig:splitterOptPseudoPot}(a) the pseudopotential is plotted in the \textit{xy}-plane. As the height of the pseudopotential minimum $z_{min}$ varies along the chip electrodes, this plot is obtained by calculating $z_{min}$ for every point along $y$ and then plotting $\Psi(x,y,z_{min}(y))$. Similarly, the pseudopotential in the \textit{zy}-plane is plotted in Fig.~\ref{fig:splitterOptPseudoPot}(b) by calculating $x_{min}(y)$ and plotting $\Psi(x_{min}(y),y,z)$. Because of fringing electric fields close to the substrate edge, the potential minimum is about $1.5\,$meV on the first $7$\,mm along the chip electrodes until quadrupole fields are fully developed leading to a field null along the guide~\cite{Hammer2014}. Further along the chip, a junction is generated in the beam splitter potential at about $y=16\,$mm. Here, an additional potential minimum path converges towards the beam splitter path from the substrate surface, as can be seen in Fig.~\ref{fig:splitterOptPseudoPot}(b).
\section{Trajectory simulations}
\label{sec: TrajSim}
We perform classical particle tracking simulations taking into account the oscillating electric field of the optimized beam splitter chip. We use the Surface Pattern package to calculate the electric field above the planar electrode structure in the gapless plane approximation. The classical particle trajectories are then obtained by numerically integrating the equation of motion for an electron in the oscillating electric field using Mathematica's built-in NDSolve function. The simulations gather 1000 particle trajectories in total that are released at the substrate edge $y=0\,$mm. More specifically, 100 rays are homogeneously distributed on a disk with a diameter of $100\,\micron$ and trajectories are released at ten different instants in time with respect to the phase of the microwave electric field. This allows us to study if the beam splitting depends on the phase of the microwave drive.

We simulate electron trajectories with microwave drive parameters of $\Omega=2\pi\cdot 1\,$GHz and $V_0=16\,$V. Fig.~\ref{fig:splitterTraj}(a) shows a top view on the simulated electron trajectories in the \textit{xy}-plane. Clearly, the electrons perform oscillations after injection into the guiding potential with a spatial period of $14$\,mm corresponding to a trap frequency of $\omega=2 \pi \cdot 50\,$MHz at an electron kinetic energy of $1.5\,$eV. In the splitting region from $y=20\,$mm to $y=30\,$mm the beam becomes symmetrically divided in the lateral $x$-direction. The color scale illustrates the initial lateral displacement of the electrons along the \textit{x}-axis. The chip electrodes are indicated in light blue. In Fig.~\ref{fig:splitterTraj}(b) the same trajectories are plotted in the vertical \textit{zy}-plane.  As can be seen, the electrons follow the beam splitter path $\bm{\gamma}(\bm{r})$ that bends down towards the substrate when approaching the beam splitter junction at $x\sim16\,$mm. Electrons released closest to the symmetry axis of the beam splitter potential [blue lines in Fig.~\ref{fig:splitterTraj}(b)] are preferentially lost from the beam splitter potential in the vertical \textit{z}-direction. This is described in detail in the main text. The simulated beam splitter output signal shown in Fig.~2(b) is obtained from the same trajectory simulations.

\begin{figure*}[htb]
\centering\includegraphics[width=0.75\textwidth]{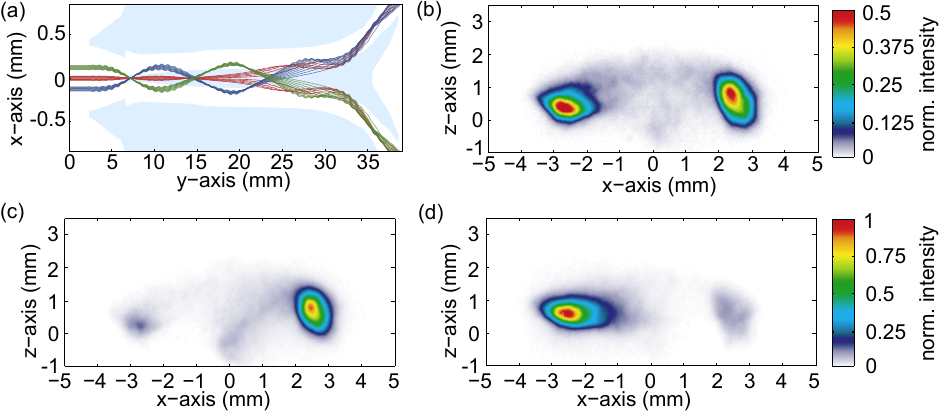}
\caption{Dependence of beam splitting on the initial position of the electron source with $E_{kin}=1\,$eV. (a) Simulated electron trajectories in the \textit{xy}-plane for three different positions of the electron source along the \textit{x}-axis. The underlying signal electrodes are indicated in light blue. (b) For a centered electron gun a symmetric beam splitting signal is measured. The measured beam splitting signals for a displaced electron gun are shown for a negative (c) and a positive (d) displacement along $x$. For a displaced electron gun [(b) and (c)] the measured count rate in one single output beam is twice the count rate in each output beam for a centered electron gun (d).}
\label{fig:splitterHorDisp}
\end{figure*}

To investigate the classical dynamics of guided electrons within the beam splitter potential we study the dependence of the beam splitting process on the initial position of the electron source by comparing particle tracking simulations with experimental measurements. We simulate electron trajectories for a centered and a displaced electron source to study the dependence of the beam splitting signal on misalignment of the electron source. Fig.~\ref{fig:splitterHorDisp}(a) shows the result of the particle tracking simulations for three different locations of the electron source along the \textit{x}-direction. The simulation as well as the corresponding measurements are performed with $E_{kin}=1\,$eV, $\Omega=2\pi\cdot1\,$GHz and $V_0=16\,$V. For a centered electron beam the trajectories (drawn in red) become symmetrically separated in the region from $y=20\,$mm to $25\,$mm. Electrons that are released at a positive (negative) \textit{x}-position end up in the output beam at positive (negative) \textit{x}-values. In contrast, for an electron source displaced about $125\,\micron$ along the positive or negative \textit{x}-direction all trajectories of the beam (drawn in green and blue) end up in the same output port at negative or positive $x$-values, respectively. The initial lateral displacement of the trajectories sets the potential energy of the transverse electron oscillation. For the initially displaced beam the potential energy of the electron oscillation is larger and electrons may cross the potential barrier in the splitting region once more compared to the centered beam.

The same behavior is found experimentally when the electron source is displaced in the \textit{x}-direction. In Fig.~\ref{fig:splitterHorDisp}(c) the electron source is displaced in the negative $x$-direction, which results in the detection of a single guided spot at positive $x$. By displacing the source in the positive $x$-direction the signal in Fig.~\ref{fig:splitterHorDisp}(d) is obtained. When the electron gun is centered we obtain a symmetric splitting, as shown in Fig.~\ref{fig:splitterHorDisp}(b). It is thus possible to modify the ratio of the electron count rate in both output beams by simply displacing the electron source. Furthermore, we find experimentally that the displacement of the electron source does not increase the signal of lost electrons. Consequently, the measured count rate in one single output beam for the displaced source corresponds to the integrated count rate of both output beams for a centered electron gun. This is reflected in the different color scale for Fig.~\ref{fig:splitterHorDisp}(b) compared to Figs.~\ref{fig:splitterHorDisp}(c),(d). For the centered beam in Fig.~\ref{fig:splitterHorDisp}(b) the color scale used spans half the intensity of the color scale used for the displaced beam in Fig.~\ref{fig:splitterHorDisp}(c) and (d).
If we move the electron source even further away along the \textit{x}-direction electron losses start to increase until no beam splitter signal is observed anymore. 

\section{Microwave design of the beam splitter chip}
\label{sec: ExpSetup}
\begin{figure}[b]
\centering\includegraphics[width=\columnwidth]{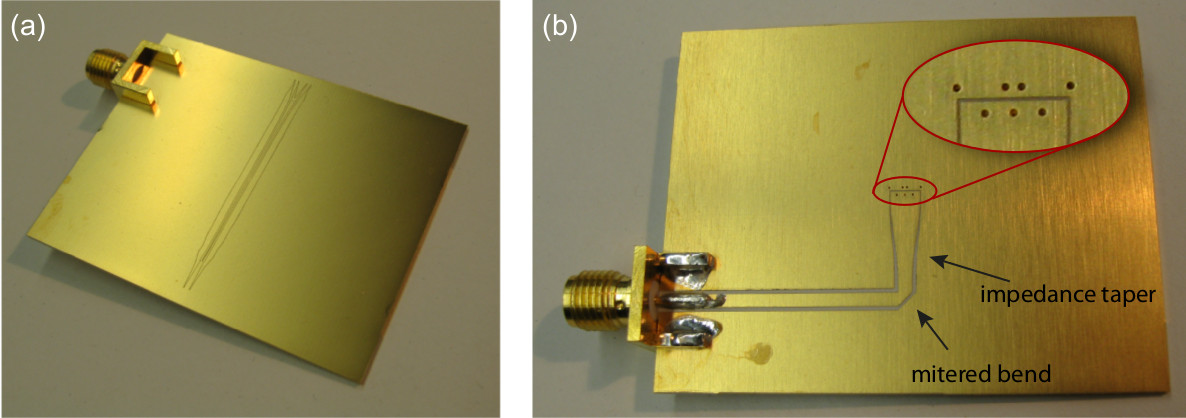}
\caption{Electron beam splitter microwave substrate. (a), Top side of the microwave substrate comprising the numerically optimized beam splitter electrodes. (b) Back side of the chip showing the microwave feeding line with the mitered bend and impedance taper for improved frequency response. The inset shows a zoom on the plated through holes with a diameter of $200\,\micron$ on the backside. They transmit the signal from the feeding line to the beam splitter electrodes.
\label{fig:splitterSubstrate} }
\end{figure}

The electron beam splitter is implemented on a planar microwave chip design that was manufactured by a commercial supplier~\cite{NoteOptiprint}. The substrate consists of a $0.76\,$mm thick microwave compatible Rogers RO4350B laminate coated with a $20\,\mu$m layer of gold-plated copper. The electrodes are defined by chemical etching of $50\,\mu$m wide gaps along the electrode contours into the metal layer. The microwave signal is delivered to the signal electrodes on the top side of the chip [shown in Fig.~\ref{fig:splitterSubstrate}(a)] by a coplanar waveguide structure on the backside of the chip [shown in Fig.~\ref{fig:splitterSubstrate}(b)], which is interconnected by laser-machined, plated through-holes with a diameter of $20\,\mu$m. 

\begin{figure*}[hbt]
  \centerline{\includegraphics[width=.75\textwidth]{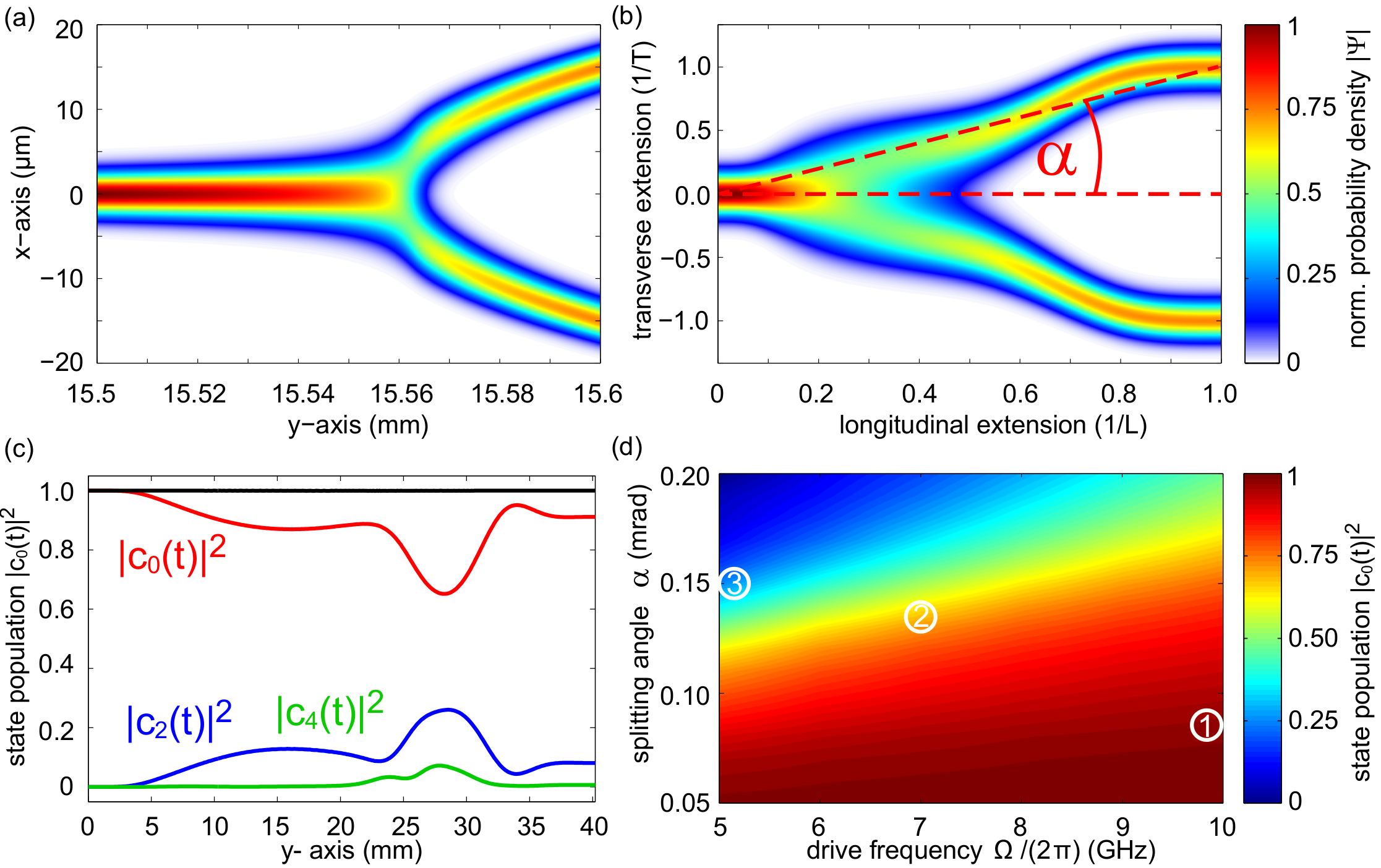}}
  \caption[guidingPot]{(Color online) Optimization of the shape of the beam splitter potential. (a) Simulation of the ground state probability density for the beam splitter potential, as experimentally realized in this work, over a length of $100\,\mu$m where the splitting of the ground state evolves. (b) Ground state probability density for the optimized beam splitter potential, which results from the adiabaticity optimization. (c) Simulated state populations of the three lowest symmetric states for the optimized beam splitter potential. For the optimized potential more than $90\%$ of the population end up in the ground state. Only symmetric states are considered as transitions occur only between states of the same parity. Details on the simulation parameters are given in the text. (d) Simulated ground state population after the splitting process as a function of the drive frequency $\Omega$ and the splitting angle $\alpha$ for $E_\mathrm{{kin}}=1\,$eV, $V_0=75\,$V. Circles are discussed in the text.
  }
  \label{fig:splitterDesignOpti}
\end{figure*}

In order to symmetrically feed all electrodes,  the feeding coplanar waveguide (cpw) structure on the back side of the chip comprises a $90^{\circ}$ mitered bend. By these means the last third of the feeding cpw is aligned parallel to the beam splitter electrodes on the top side and all electrodes are fed symmetrically. Furthermore, a triangular impedance taper was implemented in the cpw layout~\cite{Pozar2005}. This is required as the geometry of the chip electrodes typically results in a characteristic impedance of the electrode structure of $Z_0=15\,\Omega$. On the other hand, we use a standard microwave generator with a $50$-$\Omega$-matched output as well as $50$-$\Omega$ SMA connectors to transmit the microwave signal onto the guiding chip. In order to avoid reflections of the microwave signal at impedance discontinuities, the impedance taper has been implemented in the feeding cpw. Figure~\ref{fig:splitterSubstrate}(b) shows the implementation of a triangular impedance taper from $50\,\Omega$ down to $25\,\Omega$~\cite{Pozar2005}. This taper is restricted to a minimal impedance of $25\,\Omega$ because of the limited space on the backside of the chip. A taper down to $15\,\Omega$ would exceed the length of the chip.

\section{Optimization of the beam splitter potential}
\label{sec: xx}
To minimize vibrational excitations from the transverse ground state into higher energetic motional states during the beam splitting process we employ an optimization routine that systematically modifies the shape of the transverse beam splitter potential. The optimization scheme is described in detail in Ref.~\cite{Hansel2001}. The simulations take into account the one-dimensional beam splitter potential along the $x$-direction, as this is the dimension where the splitting arises. Furthermore, only the time-averaged pseudopotential is considered.

To find the eigenfunctions of the one-dimensional Schrödinger equation in the transverse $x$-direction, $\left\{-\frac{\hbar^2}{2m}\frac{\partial^2}{\partial x^2} +V(x)\right\}\psi(x) = E \psi(x)$, we only look for solutions in a region of length $X$ around the center of the guiding potential. $X$ must be significantly larger than the extent of the eigenfunctions of interest. We then expand the Hamilitonian in the basis of standing waves with wavelengths of $2X/n$ up to a finite order $n= 500$. By diagonalizing the resulting matrix using built-in Matlab functions, we obtain a good approximation of the eigenfunctions and eigenenergies of the Hamiltonian for orders $\ll n$. This numerical procedure is described in detail in Ref.~\cite{Jelic2012}.

Figure~\ref{fig:splitterDesignOpti}(a) shows the simulated ground state probability density over a length $L=100\,\mu$m along the $y$-direction where the splitting of the quantum ground state arises. In the adiabatic limit an electron wave packet, initially prepared in the ground state, continuously adapts its wavefunction to the ground state probability density when propagating along the beam splitter potential. If adiabaticity of the splitting process cannot be assured non-adiabatic propagation of the electron wave packet within the splitting potential manifests itself by conversion of longitudinal momentum into the transverse degree of freedom, thereby exciting the electron wave into a higher energy state of the transverse confining potential.

These transverse vibrational excitations depend critically on the precise shape of the beam splitter potential. To find its optimum shape we follow an optimization routine that was initially developed to achieve fast and adiabatic splitting of cold atomic clouds in an atom chip magnetic trap~\cite{Hansel2001b}. The optimization routine parametrizes the beam splitter potential along the longitudinal extension, effectively deforming the potential along $y$ by stretching it locally. As a result, a beam splitter potential is obtained that increases the adiabaticity of the wave propagation and reduces vibrational excitations from the ground state into excited states. Fig.~\ref{fig:splitterDesignOpti}(b) shows the simulated quantum ground state probability density for the improved beam splitter potential obtained from the optimization. As expected, a smooth transition into the split-up paths by means of a small splitting angle $\alpha$ is required and obtained from the optimization. Here $\alpha$ is defined as the ratio between the transverse extension $T$, defined as half the separation length at the output, and the length $L$ of the beam splitter.

We study vibrational excitations that arise during the beam splitting process by calculating the state population $\left|c_i(t)\right|^2$ for the ten lowest symmetric quantum states by solving the time-dependent Schrödinger equation (see equation (9) in Ref.~\cite{Hansel2001b}). Only symmetric states are considered, as transitions occur only between states of the same parity due to the planar symmetry of the splitting potential. We assume that initially only the ground state is populated. The solid lines in Fig.~\ref{fig:splitterDesignOpti}(c) show the temporal evolution of the state populations (i=0,\,2,\,4) for $E_{\mathrm{kin}}=1\,$eV along the optimized beam splitter potential. We find that with $90\%$ probability an electron wave packet remains in the ground state after the splitting process, even though during splitting the excited state population may transiently reach values up to $26\%$. The simulation is performed with $\Omega=2 \pi \cdot 8\,$GHz~\cite{NoteElectricallyLong,Hoffrogge2011b}. This corresponds to an eightfold increased trap frequency $\omega$ with respect to the measurements presented in this paper ($\omega\propto\Omega$ for constant $q$). A constant stability parameter $q=0.15$ is assured by increasing the voltage amplitude to $V_0=75\,$V and scaling the beam splitter potential in the transverse dimension by a factor $3.7$, which results in a relative reduction of the trap height $R'_0=R_0/3.7$, which is then on the order of $100\,\micron$. Furthermore, the section of the beam splitter potential that underlies the probability density simulation in Fig.~\ref{fig:splitterDesignOpti}(c) is scaled longitudinally to a length $L=40\,$mm. The scaling of the beam splitter potential results in a splitting angle $\alpha=0.1\,$mrad and a smaller beam separation of $\sim 8\,\micron$ at the end of the beam splitter chip as compared to a separation of $2\,$mm in the experiments described in the main text. A discussion on the technical realization of these parameters is beyond the scope of this letter. 

It is instructive to specify how excitations from the ground state scale with $\Omega$ (and hence $\omega$) and $\alpha$.
Figure~\ref{fig:splitterDesignOpti}(d) shows the ground state population probability $\left|c_0\right|^2$ after the splitting for varying $\Omega$ and $\alpha$.
Clearly, excitations are reduced for small splitting angles $\alpha$ and large $\Omega$.
For $\alpha=0.05\,$mrad and $\Omega=2 \pi \cdot 10\,$GHz we find that more than $95\%$ of the ground state population remains in its state during the entire beam splitting process giving rise to a nearly adiabatic trajectory.
\section{Matrix representation of the microwave beam splitter}
In order to investigate the quantum dynamics of an electron wave in the microwave beam splitter for electrons and to compare it to a typical $50/50$-beam splitter as used in light optics, we have carried out a one-dimensional wave packet simulation in Matlab using the split step method~\cite{Fleck1976,Feit1982}. As described in the main text of the manuscript, we extend the optimized beam splitter potential discussed above to an X-shaped one with two input and two output ports. This could be done by using two splitters in sequence or by placing an electron mirror at the single port of the Y-shaped splitter. In these simulations we numerically solve the time-dependent Schrödinger equation taking into account the optimized beam splitter potential and assuming a free particle with $E_{\mathrm{kin}}=1\,$eV along the $y$-direction.
\label{sec: xxx}
\begin{figure*}[tb]
  \centerline{\includegraphics[width=\textwidth]{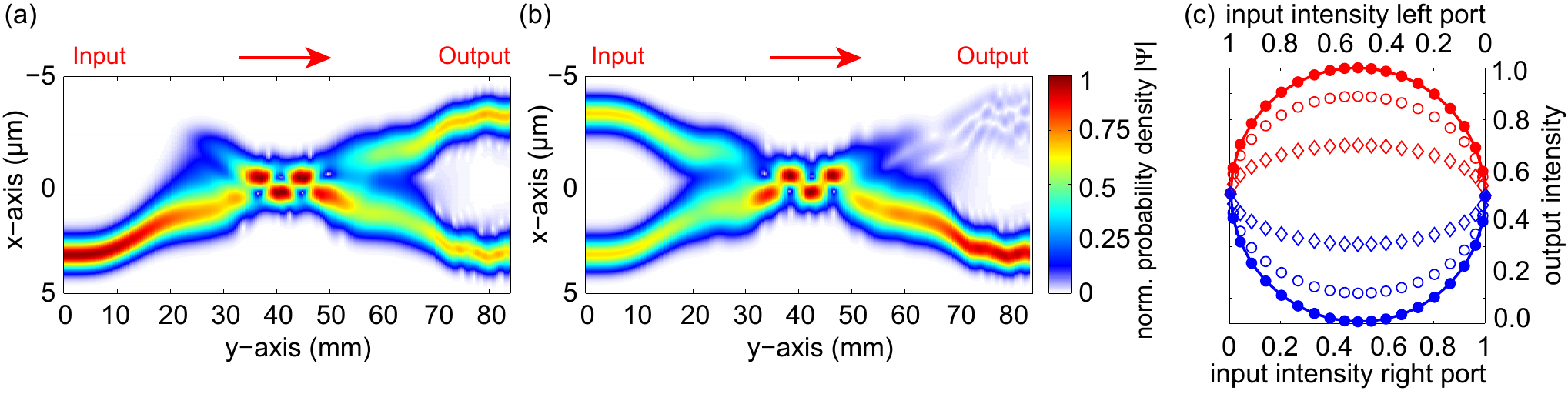}}
  \caption[guidingPot]{(Color online) Quantum matter-wave simulations of a $50/50$-\,microwave beam splitter. (a) Injection of the electron wave into the ground state of the input port $\rig$ results in a balanced output with equal intensities in $\rig$ and $\lef$. (b) For a balanced input state interference gives rise to a bright port ($\rig$) and a dark port ($\lef$). (c) Simulated output intensities as a function of the beam splitter input. The red markers correspond to the output intensity $r'^2$ in state $\rig$ and the blue markers to the intensities $l'^2$ in state $\lef$. The marker symbols represent different simulation parameters $\alpha$ and $\Omega$, as described in the text.}
  \label{fig:BSanalogy}
\end{figure*}

Figure~\ref{fig:BSanalogy}(a) shows the action of the beam splitter when an electron wave is injected into the ground state of the lower input port, labeled as $\rig$ in the main text. The simulation parameters are $\alpha=0.08\,$mrad, $\Omega=2\pi\cdot 10$\,GHz and $V_0=75\,$V. This localized input state corresponds to a superposition $\rig=(\eigone-\eigtwo)/\sqrt{2}$ of the energy eigenstates of the beam splitter potential. As a result, the electron wave performs an oscillation in the central region from $y=30$\,mm to $y=50$\,mm, where the potential is given by a single well. By tuning the length of the beam splitter in this center part the splitting ratio can be adjusted. Here, the intensity ratio between $\lef$ and $\rig$ in the output of the beam splitter can be $50\%$. Vibrational excitations into higher energetic states manifest themselves as small distortions of the electron wave in both output arms. In contrast, in Fig.~\ref{fig:BSanalogy}(b), for a balanced input in $\lef$ and $\rig$, interference results in a dark and a bright port at the output of the beam splitter. The small deviation from an ideal dark port, as visible by the almost negligible intensity in $\lef$, results from higher energetic states that become excited during the splitting process. The phase of the wave amplitude in input state $\lef$ is shifted by $\pi/2$ with respect to the wave amplitude in $\rig$ in order to obtain the desired splitting ratio. As described in the main text, the effect of an \textit{amplitude} beam splitter can be described as a multiplication of the input states $\lef=\bigl(\begin{smallmatrix} 1 \\ 0 \end{smallmatrix} \bigr)$ and $\rig=\bigl(\begin{smallmatrix} 0 \\ 1 \end{smallmatrix} \bigr)$ with a unitary matrix $\bigl(\begin{smallmatrix}
 l' \\ r'
\end{smallmatrix}\bigr)
= B
\bigl(\begin{smallmatrix}
 l \\ r
\end{smallmatrix}\bigr)$. Fig.~\ref{fig:BSanalogy}(c) shows the beam splitter output intensities, where blue corresponds to $l'^2$ and red to $r'^2$, as a function of the input intensity ratio. The filled circles correspond to the simulation parameters given above and also used in Fig.~\ref{fig:BSanalogy}(a),(b). Here, vibrational excitations can be almost neglected [see marker $\raisebox{.5pt}{\textcircled{\raisebox{-.9pt} {1}}} $ in Fig.~\ref{fig:splitterDesignOpti}(c)]. The simulated output intensities perfectly match the analytically calculated output intensities, as given by the matrix $B=\bigl(\begin{smallmatrix} \mathrm{cos}(\theta)~-\mathrm{sin}(\theta)\\ \mathrm{sin}(\theta)~\mathrm{cos}(\theta) \end{smallmatrix} \bigr)$ with $\theta=\pi/4$, which fully describes the action of the beam splitter. The solid lines in Fig.~\ref{fig:BSanalogy}(c) correspond to the analytically calculated output intensities obtained from a matrix multiplication with $B$. The open circles in Fig.~\ref{fig:BSanalogy}(c) correspond to different simulation parameters $\alpha=0.13\,$mrad and $\Omega=2\pi\cdot 7$\,GHz. As can be seen, for these simulation parameters the shape of the output intensities turns into an ellipse and not all splitting ratios can be realized any more. This can be explained by vibrational excitations allowing only $72\%$ percent of the population to remain in the ground state [see marker $\raisebox{.5pt}{\textcircled{\raisebox{-.9pt} {2}}} $ in Fig.~\ref{fig:splitterDesignOpti}(c)]. As a result, the excitation of higher energetic states reduces the contrast in the achievable splitting ratios and causes a deviation from the matrix representation described above. This effect becomes even more apparent in the simulation result with $\alpha=0.15\,$mrad and $\Omega=2\pi\cdot 5$\,GHz [see marker $\raisebox{.5pt}{\textcircled{\raisebox{-.9pt} {3}}} $ in Fig.~\ref{fig:splitterDesignOpti}(c)], which is indicated by the diamonds forming an even more elongated ellipse.


\begin{thebibliography}{10}
\expandafter\ifx\csname url\endcsname\relax
  \def\url#1{{\tt #1}}\fi
\expandafter\ifx\csname urlprefix\endcsname\relax\def\urlprefix{URL }\fi

\bibitem{Mandel1995}
L.~Mandel and E.~Wolf.
\newblock {\em Optical Coherence and Quantum Optics\/} (Cambridge University
  Press, 1995).

\bibitem{Bloch2008}
I.~Bloch, J.~Dalibard, and W.~Zwerger.
\newblock Rev.~Mod.~Phys. {\bf 80}, 885 (2008).

\bibitem{Rauch2000}
H.~Rauch and S.~Werner.
\newblock {\em Neutron Interferometry\/}.
\newblock (Oxford University Press, 2000).


\bibitem{Juffmann2012}
Th.~Juffmann, {\em et~al.\/}.
\newblock Nat.~Nanotech. {\bf 7}, 297 (2012).

\bibitem{Davisson1927}
C.~Davisson and L.~H. Germer.
\newblock Phys.~Rev. {\bf 30}, 705 (1927).

\bibitem{Boersch1943}
H.~Boersch.
\newblock Physik Z. {\bf 44}, 202 (1943).

\bibitem{Marton1953}
L.~Marton, J.~A. Simpson, and J.~A. Suddeth.
\newblock Phys. Rev. {\bf 90}, 490 (1953).

\bibitem{Tonomura1989}
A.~Tonomura, J.~Endo, T.~Matsuda, T.~Kawasaki, and H.~Ezawa.
\newblock Am. J. Phys. {\bf 57}, 117 (1989).

\bibitem{Tonomura1986}
A.~Tonomura, {\em et~al.\/}.
\newblock Phys.~Rev.~Lett. {\bf 56}, 792 (1986).

\bibitem{Hasselbach2010}
F.~Hasselbach.
\newblock Rep.~Progr.~Phys. {\bf 73}, 016101 (2010).

\bibitem{Mollenstedt1955}
G.~M{\"o}llenstedt and H.~D{\"u}ker.
\newblock Naturw. {\bf 42}, 41 (1955).

\bibitem{Gabor1948}
D.~Gabor.
\newblock Nature {\bf 161}, 777 (1948).

\bibitem{Tonomura1999}
A.~Tonomura.
\newblock {\em Electron Holography\/}.
\newblock (Springer, Heidelberg, 1999).

\bibitem{Germann2010}
M.~Germann, T.~Latychevskaia, C.~Escher, and H.-W. Fink.
\newblock Phys.~Rev.~Lett. {\bf 104}, 095501 (2010).

\bibitem{Hoffrogge2011}
J.~Hoffrogge, R.~Fr\"ohlich, M.~A. Kasevich, and P.~Hommelhoff.
\newblock Phys.~Rev.~Lett. {\bf 106}, 193001 (2011).

\bibitem{Amini2010}
J.~M. Amini, {\em et~al.\/}.
\newblock New~J.~Phys. {\bf 12}, 033031 (2010).

\bibitem{Moehring2011}
D.~L. Moehring, {\em et~al.\/}.
\newblock New~J.~Phys. {\bf 13}, 075018 (2011).

\bibitem{Shu2014}
G.~Shu, {\em et~al.\/}.
\newblock Phys.~Rev.~A {\bf 89}, 062308 (2014).

\bibitem{Pearson2006}
C.~E. Pearson, {\em et~al.\/}.
\newblock Phys.~Rev.~A {\bf 73}, 032307 (2006).

\bibitem{Wright2013}
K.~Wright, {\em et~al.\/}.
\newblock New~J.~Phys. {\bf 15}, 033004 (2013).

\bibitem{Hensinger2006}
W.~K. Hensinger, {\em et~al.\/}.
\newblock Appl.~Phys.~Lett. {\bf 88} (2006).

\bibitem{Brown2011}
K.~R. Brown, {\em et~al.\/}.
\newblock Nature {\bf 471}, 196 (2011).

\bibitem{Harlander2011}
M.~Harlander, R.~Lechner, M.~Brownnutt, R.~Blatt, and W.~H\"ansel.
\newblock Nature {\bf 471}, 200 (2011).

\bibitem{Major2005}
F.~G. Major, V.~N. Gheorghe, and G.~Werth.
\newblock {\em Charged Particle Traps\/} (Springer, Heidelberg, 2005).

\bibitem{Paul1990}
W.~Paul.
\newblock Rev.~Mod.~Phys. {\bf 62}, 531 (1990).

\bibitem{Wesenberg2009}
J.~H. Wesenberg.
\newblock Phys.~Rev.~A {\bf 79}, 013416 (2009).

\bibitem{Hommelhoff2014}
P.~Hommelhoff and J.~Hammer.
\newblock Patent pending (2014).

\bibitem{Wesenberg2008}
J.~H. Wesenberg.
\newblock Phys.~Rev.~A {\bf 78}, 063410 (2008).

\bibitem{NoteSurfacePattern}
See http://atom.physik.unibas.ch/people/romanschmied/ \mbox{code/SurfacePattern.php.}

\bibitem{Schmied2010}
R.~Schmied.
\newblock New~J.~Phys. {\bf 12}, 023038 (2010).

\bibitem{Schmied2009}
R.~Schmied, J.~H. Wesenberg, and D.~Leibfried.
\newblock Phys.~Rev.~Lett. {\bf 102}, 233002 (2009).

\bibitem{NoteElectricallyShort}
A standing microwave signal is established as their wavelength $\lambda=200\,$mm is much larger than the electrode length $L=38\,$mm.

\bibitem{Erdman1982}
P.~W. Erdman and E.~C. Zipf.
\newblock Rev.~Sci.~Instr. {\bf 53}, 225 (1982).

\bibitem{NoteMCP}
{\em \rm{Photonis, Model: APD 2 PS 40/12/10/12 46:1 P20}\/}.

\bibitem{NoteCCD}
{\em \rm{The Imaging Source, Model: DMK 41AU02}\/}.

\bibitem{Hammer2014}
J.~Hammer, J.~Hoffrogge, S.~Heinrich, and P.~Hommelhoff.
\newblock Phys. Rev. Applied {\bf 2}, 044015 (2014).

\bibitem{Andersson2002}
E.~Andersson, {\em et~al.\/}.
\newblock Phys.~Rev.~Lett. {\bf 88}, 100401 (2002).

\bibitem{Zeilinger1981}
A.~Zeilinger.
\newblock Am. J. of Physics {\bf 49}, 882 (1981).

\bibitem{Schleich2005}
W.~P. Schleich.
\newblock {\em Quantum Optics in Phase Space\/} (Wiley-VCH, Berlin, 2005).

\bibitem{Born1999}
M.~Born and E.~Wolf.
\newblock {\em Principles of Optics\/} (Cambridge University Press, 1999), 7
  edn.

\bibitem{Putnam2009}
W.~P. Putnam and M.~F. Yanik.
\newblock Phys.~Rev.~A {\bf 80}, 040902 (2009).

\bibitem{Thomas2014}
S.~Thomas, C.~Kohstall, P.~Kruit, and P.~Hommelhoff.
\newblock Phys. Rev. A {\bf 90}, 053840 (2014).

\bibitem{NoteOptiprint}
{\em \rm{{Optiprint AG}, Auerstrasse 37, CH-9442 Berneck, Switzerland,
  www.optiprint.ch}\/}.

\bibitem{Pozar2005}
D.~M. Pozar.
\newblock {\em Microwave Engineering\/} (John Wiley and Sons, 2005), 3 edn.

\bibitem{Hansel2001}
W.~H{\"a}nsel, P.~Hommelhoff, T.~W. H{\"a}nsch, and J.~Reichel.
\newblock Nature {\bf 413}, 498 (2001).

\bibitem{Jelic2012}
V.~Jelic and F.~Marsiglio.
\newblock Eur. J. Phys. {\bf 33}, 1651 (2012).

\bibitem{Hansel2001b}
W.~H{\"a}nsel, J.~Reichel, P.~Hommelhoff, and T.~W. H{\"a}nsch.
\newblock Phys.~Rev.~A {\bf 64}, 063607 (2001).

\bibitem{NoteElectricallyLong}
Here traveling microwave signals have to be considered, as the on-chip
  microwave wavelength $\lambda$ becomes smaller than the longitudinal
  electrode length $L$.

\bibitem{Hoffrogge2011b}
J.~Hoffrogge and P.~Hommelhoff.
\newblock New J. Phys. {\bf 13}, 095012 (2011).

\bibitem{Fleck1976}
J.~Fleck, J.A., J.~Morris, and M.~Feit.
\newblock Applied Physics {\bf 10}, 129 (1976).

\bibitem{Feit1982}
M.~Feit, J.~Fleck, and A.~Steiger.
\newblock Journal of Computational Physics {\bf 47}, 412  (1982).

\end{thebibliography}
\end{document}